\documentclass[A4,11pt]{article}
\usepackage{amsmath,amssymb}
\usepackage{euscript}

\def\a{{\alpha}}
\def\b{{\beta}}

\def\0{\nonumber}

\def\log{{\rm log}}

\newcommand\ee{\end{eqnarray}}      
\newcommand\be{\begin{eqnarray}}
\newcommand\ba{\begin{array}}           
\newcommand\ea{\end{array}}
\newcommand\eeq{\end{equation}}     
\newcommand\beq{\begin{equation}}

\title{TFD Extension of the Open String Field Theory }

\author{M. Botta Cantcheff$^{(a)}$\footnote{e-mail:
bottac@cern.ch, botta@fisica.unlp.edu.ar} , R. J. Scherer Santos$^{(b)}$\footnote{e-mail: rscherer@ufrrj.br}}

\begin{document}

\maketitle

\begin{center}
$(a)$ {\it IFLP-CONICET and Departamento de F\'{\i}sica\\ Facultad de Ciencias Exactas, Universidad Nacional de La Plata\\
CC 67, 1900,  La Plata, Argentina}
\end{center}

\begin{center}
$(b)$ {\it Departamento de F\'{i}sica, Universidade Federal Rural do Rio de Janeiro, \\
BR 465-07, 23890-971, Serop\'{e}dica, RJ, Brazil}
\end{center}

\abstract{
We study the application of the rules of Thermo Field Dynamics (TFD) to the covariant formulation of Open String Field Theory (OSFT). We extend the states space and fields according to the duplication rules of TFD and
construct the corresponding classical action. The result is interpreted as a theory whose fields would encode the statistical information of open strings.

The physical spectrum of the free theory is studied through the cohomology of the extended BRST charge, and, as a result, we get new fields in the spectrum emerging by virtue of the quantum entanglement and, noticeably, it presents degrees of freedom that could be identified as those of closed strings. We also show, however, that
their appearing in the action is directly related to the choice of the inner product in the extended algebra, so that different sectors of fields could be
 eliminated from the theory by choosing that product conveniently.

Finally, we study the extension of the three-vertex interaction and provide a simple prescription for it whose results at tree-level agree with those of the conventional theory.
}

\section{Introduction}

Open String Field Theory (OSFT) is deemed to be the right arena where to study nonperturbative aspects of string theory and, possibly, even a way to define it nonperturbatively \cite{Witten1,GJ1,GJ2,Gaume,CST,Ohta,leclair1,leclair2}. A lot of progress has been made since its appearing in the covariant formulation \cite{Witten1}, and it has been proven to be the right way to address some important issues such as, for example, the condensation of the tachyon field \cite{Ohmori,TZ,RSZ,tope,BMP1,BMP2,RSZ3,RSZ2}. In past years, a lot of progress has been made concerning the finding of classical solutions of OSFT, showing its capability of studying the vacua of string theory and their physical interpretation. Particularly, some of these are to be associated with D-branes and/or their decaying \cite{BMST1,MST,Schn2,Schn3,Oka,FK1,FK2,RZ,EScn,KORZ,KOSoler,BMTolla,BGTolla,BGTolla1,EMac,EKroyter}.


On the other hand, Thermo Field Dynamics (TFD), developed by Takahashi and Umezawa \cite{ume2,ume4,rev2,ume1,kha2,kha3,ume12},
is a real time approach to
quantum field theory at finite temperature \cite{kob1,leb1} where
an identical but fictitious copy of the system is properly introduced.
 The state space is the tensor product of two copies of the original Hilbert space, and
  the thermodynamical information of a quantum system is encoded in a fundamental state in this space instead of the density matrix.
More generally, the statistical average of an operator
 ${\cal O}$ can be defined as its expectation value
in a certain (ground) state in the extended Hilbert space:
\begin{equation}{\mbox Tr}[{\cal O} \, \rho]
= \left\langle \Omega  \left| {\cal O}
\right| \Omega \right\rangle
\label{tr1}
\end{equation}
for any density matrix $\rho$.
 In particular, at thermal equilibrium, the density matrix $\rho_\beta = e^{-\b H}/ Z $ corresponds to the thermal ground state
\be |\Omega(\b) \rangle = Z^{-1 /2} \sum_n  e^{ -\b E_n /2 } | n \rangle
|\tilde{n} \rangle \, \,\, \in \, {\cal H} \otimes \widetilde{{\cal H}}\,,\ee
where $| n ; \tilde{n} \rangle$ denotes the
$n^{th}$ energy eigenvalue of the two systems.
In this way, one could describe any mixed state by a pure (but entangled) state in the extended Hilbert space.

The purpose of this paper is to study in depth how this extension can be constructed in the context of OSFT. In this sense, our construction here is nothing but a theory for the general mixed/entangled states of open strings (eq. (\ref{tr1})), rather than for thermal states. In fact this is the necessary step, previous to introduce thermal equilibrium and temperature, which shall be realized in a forthcoming work.

 The specific goal of describing thermal OSFT using TFD was first achieved by Leblanc \cite{leblanc} shortly after the appearing of Witten's formulation of OSFT \cite{Witten1}. It was done by decomposing the open string field in an infinity of ordinary point-like quantum fields and using the standard rules of TFD to compute the thermal correlation functions. Because of this, the application of the TFD rules to \emph{non-local} objects as the string field has been lacking,  and the duplication TFD principles have not been yet incorporated to the axiomatic structure of OSFT. In fact, the main motivation of this work is to study the implications of the TFD extension on the modern formulation of OSFT, in view of these unexplored aspects\footnote{In another context, properties of first quantized strings and D-branes at finite temperature have already been studied \cite{vazquez}, and the idea of using TFD to study D-branes at finite temperature came up in \cite{IVV,AGV2,AEG,AGV3,AGV4,AGV5,braneTFD}.}.


Other aspect to have in mind, apparently unrelated to the discussion above, is the connection with closed strings. It has been suggested by A. Sen that open string theory might be able to describe some (if not all) of the closed string physics, at least in a background of D--branes \cite{Sen,Sen1}. In this sense
open string field theory should be a privileged ground to check this idea. Open string field theory is of course formulated in terms of open strings degrees of freedom, but there is ample evidence that tachyon condensation leads to a new vacuum, and that this new vacuum is the closed string one. If, as expected, the gravitational interaction can be described as emerging from OSFT, considered fundamental, then backgrounds containing black objects that have thermal properties should be described by the own OSFT in some sense. Thus, a study of the thermal/statistical properties of OSFT is required. We are going to show that a TFD extension of OSFT indeed captures some features of the gravitational interaction and closed strings.

This article is organised as follows. In Section 2, we briefly review the OSFT and present some standard formulas. In Section 3, we introduce the basic rules of the TFD formalism and, in order to construct the extended string fields, we extend the open string vacuum accordingly. In Section 4, we study how the TFD extension of the free OSFT can be implemented and, in order to define the kinetic term, we discuss the issue of how to define an appropriate inner product in the extended space of fields. In Section 5, we construct the more general ground states of the theory that would encode the statistical information (as eq. (\ref{tr1})), and show that their field content is coincident with the low energy spectrum of closed strings. Moreover, it is argued that, for the proper choice of the inner product, the present theory agrees with the free action of closed string field theory.
In Section 6, we investigate the spectrum of physical fields (for the lower levels) by studying the cohomology of the extended BRST charge operator on the extended space. The extended OSFT action is wrote down in Section 7 and it is argued that the explicit dependence on the more general inner product provides a mechanism to eliminate many sectors of fields. In Section 8, we propose the simplest prescription for the extended vertex (star product) in order to reproduce the conventional OSFT dynamics at tree level. Concluding remarks are collected in Section 9 where we stress that this theory is equipped with a natural definition of entropy.

\section{Open String Field Theory: Preliminaries. }

The main object in this theory is the string field $|\Phi \rangle$, which is an element of a graded algebra $\cal{A}$. In this algebra, a star product is defined $\star: \cal{A} \otimes \cal{A} \rightarrow \cal{A}$. This product is additive with respect to the degree. There are also a BRST operator $Q$ of degree $1$ and an integral operation which takes the string field to a complex number. These elements are required to satisfy a set of axioms:

\begin{eqnarray}
&{\bf i)}& ~~Q^{2} |\Phi\rangle = 0 ~~;~~ \forall |\Phi\rangle \in \cal{A} \nonumber \\
&{\bf ii)}& ~~ \int Q |\Phi \rangle = 0 ~~;~~ \forall |\Phi\rangle \in \cal{A} \nonumber \\
&{\bf iii)}& ~~ Q (|\Phi\rangle \star |\Psi\rangle) = (Q|\Phi\rangle) \star |\Psi\rangle + (-1)^{\Phi} |\Phi\rangle \star (Q |\Psi\rangle) ~~;~~ \forall |\Phi\rangle , |\Psi\rangle \in \cal{A} \label{SFTaxioms}\\
&{\bf iv)}& ~~ \int |\Phi\rangle \star |\Psi\rangle = (-1)^{\Phi \Psi} \int |\Psi\rangle \star |\Phi\rangle ~~;~~ \forall |\Phi\rangle , |\Psi\rangle \in \cal{A} \nonumber \\
&{\bf v)}& ~~ (|\Phi\rangle \star |\Psi\rangle) \star |\Xi\rangle = |\Phi\rangle \star (|\Psi\rangle \star |\Xi\rangle) ~~;~~ \forall |\Phi\rangle , |\Psi\rangle , |\Xi\rangle \in \cal{A} \nonumber
\end{eqnarray}

An action is then postulated

\begin{equation}
S = \frac{1}{2} \langle \Phi , Q\Phi \rangle + \frac{g}{3}  \langle \Phi , \Phi \star \Phi \rangle ,
\label{SFTaction}
\end{equation}
where $g$ is the open string coupling constant. Once the axioms (\ref{SFTaxioms}) are satisfied, this action is invariant under the gauge transformation

\begin{equation}
\delta|\Phi\rangle = Q |\Lambda\rangle + |\Phi\rangle \star |\Lambda\rangle - |\Lambda\rangle \star |\Phi\rangle,
\label{gaugetransf}
\end{equation}
where $|\Lambda\rangle \in \cal{A}$ is a gauge parameter with degree $0$.

If we take the string field as a functional of the matter and ghost fields that describe a string in a 26-dimensional space-time, the BRST operator $Q$ to be the BRST operator $Q_{B}$ of the open string and the degree of the algebra to be associated with the ghost number of the string field\footnote{In this paper, we denote the ghost number of a string field as $gh(\Phi)$, and it is computed by the usual rule: the number of ghosts (c) minus the number of anti-ghosts (b).}, it has been shown that all these axioms and structure are satisfied. The string field can then be expanded in terms of the Fock space states of the open string with their coefficients being space-time fields:

\begin{equation}
|\Phi \rangle = \int \frac{d^{26} k}{(2\pi)^{26}} \left[  t(k) + A_{\mu}(k) \alpha^{\mu}_{-1} + \cdots \right] |\Omega \rangle,
\label{stringfield}
\end{equation}
where $|\Omega \rangle = c_{1} |0;k \rangle$ is the Fock space vacuum. The star product is naturally defined as the gluing of the right half of one string to the left half of the other producing a third string, defining in this way how strings interact. Finally, the integration operation is performed by gluing the left and right halves of the string.

The brackets in (\ref{SFTaction}) are defined by

\begin{equation}
\langle \Phi | \Psi \rangle \equiv \langle bpz(\Phi) | \Psi \rangle
\label{inproduct}
\end{equation}
where $bpz$ operation is defined as follows. For a primary field $\phi(z)$, taking the $bpz$ means transforming this field by $I(z) = - 1/z$. In terms of the modes\footnote{A primary field of conformal weight $h$ has an expansion in terms of its modes as $\phi(z) = \sum_{n=-\infty}^{\infty} \frac{\phi_{n}}{z^{n+h}}$.} of the primary field, the $bpz$ operation means

\begin{equation}
{\rm bpz}(\phi_{n}) = (-1)^{n+h} \phi_{-n}
\label{bpz}
\end{equation}
The bracket (\ref{inproduct}) is also written, in terms of the world-sheet conformal field theory, as an amplitude to be computed on the unitary disk in the complex plane, that is

\begin{equation}
\langle \Phi | \Psi \rangle \equiv \langle I \circ \Phi(0) \Psi(0) \rangle_{Disk}
\label{inproductCFT}
\end{equation}

The interaction term in (\ref{SFTaction}) represents the gluing of three strings and is defined in terms of an amplitude as

\begin{equation}
\langle \Phi, \Psi \star \Xi \rangle \equiv \langle f_{1} \circ \Phi(0) f_{2} \circ \Psi(0) f_{3} \circ \Xi(0) \rangle_{Disk}
\label{intterm}
\end{equation}
where $f_{i}$ are the functions that map each of the upper-half disks of each string, given by coordinates $\xi_{i}$, to the unit disk in the $w$ complex plane:

\begin{equation}
f_{1}(\xi_{1}) = e^{\frac{2 \pi i}{3}} \left( \frac{1 +i \xi_{1}}{1 - i \xi_{1}} \right)^{\frac{2}{3}} ~;~ f_{2}(\xi_{2}) = \left( \frac{1 +i \xi_{2}}{1 - i \xi_{2}} \right)^{\frac{2}{3}} ~;~ f_{3}(\xi_{3}) = e^{-\frac{2 \pi i}{3}} \left( \frac{1 +i \xi_{3}}{1 - i \xi_{3}} \right)^{\frac{2}{3}}
\end{equation}

One object that will be useful in the following is the reflector state. It is a state that lives in $\cal{H}^{*} \otimes \cal{H}^{*}$ and is defined, in connection to the kinetic term of the action, as

\begin{equation}
\langle \Phi | \Psi \rangle =  \langle R_{12}| |\Phi \rangle_{1} |\Psi \rangle_{2}
\label{defreflector}
\end{equation}
where the subscripts $1$ and $2$ refer to the Hilbert spaces of the first and second string, respectively. One can then obtain an expression for the reflector in terms of the oscillator modes of the matter and ghost fields \cite{GJ1,GJ2}:

\begin{eqnarray}
\langle R_{12}| &=& \int \frac{d^{26}k}{(2\pi)^{26}} (_{1} \langle 0;k| c_{-1} \otimes ~_{2} \langle 0;-k|  c_{-1}) (c_{0}^{(1)} + c_{0}^{(2)}) \nonumber \\
&\times& {\rm exp} \left( -\sum_{n=1}^{\infty} (-1)^{n} \left[ a_{n}^{\mu (1)} a_{\mu , n}^{(2)} + c_{n}^{(1)} b_{n}^{(2)} + c_{n}^{(2)} b_{n}^{(1)} \right] \right)
\label{reflectorO}
\end{eqnarray}
where the oscillator modes $a^{\mu}_{n}$ (matter) and $c_{n}$, $b_{n}$ (ghosts) obey the known algebras

\begin{equation}
[a^{\mu}_{n}, a^{\nu \dagger}_{m}] = \delta_{m,n} \eta^{\mu \nu} ~~;~~ \{c_{n} , b_{m} \} = \delta_{m+n}
\end{equation}
where $a^{\mu}_{n} = \alpha^{\mu}_{n}/ \sqrt{n}$ for $n > 0$.

Also the interaction term of the action can be written is terms of a state called the three-string vertex. The three-string vertex is defined as a state $\langle V_{123}| \in \cal{H}^{*} \otimes \cal{H}^{*} \otimes \cal{H}^{*}$ such that

\begin{equation}
\langle \Phi, \Psi \star \Xi \rangle \equiv \langle V_{123} | |\Phi\rangle_{1} |\Psi\rangle_{2} |\Xi\rangle_{3},
\label{3stvertdef}
\end{equation}
where the subscripts $1,2$ and $3$ refer to the Hilbert spaces of the first, second and third string respectively. It was shown that this vertex can be written (as a ket) as

\begin{eqnarray}
| V_{123} \rangle &=& {\cal N} \int \frac{d^{26}k^{(1)}}{(2\pi)^{26}} \frac{d^{26}k^{(2)}}{(2\pi)^{26}}  \frac{d^{26}k^{(3)}}{(2\pi)^{26}} \nonumber \\
&\times& {\rm exp} \left( \sum_{r,s = 1}^{3} \sum_{m,n} -\frac{1}{2} a_{m}^{(r)} V^{rs}_{mn} a_{n}^{(s)} - a_{m}^{(r)} V^{rs}_{m0} k^{(s)} - \frac{1}{2} k^{(r)} N^{rs}_{00} k^{(s)} - c_{m}^{(r)} X^{rs}_{mn} b_{n}^{(s)} \right) \nonumber \\
&\times& \delta(k^{(1)} + k^{(2)} + k^{(3)}) c_{0}^{(1)} c_{0}^{(2)} c_{0}^{(3)} \left( |\Omega \rangle_{1} \otimes |\Omega \rangle_{2} \otimes |\Omega \rangle_{3} \right),
\label{3vertex1}
\end{eqnarray}
where the constant coefficients $V^{rs}_{mn}$, $V^{rs}_{m0}$, $V^{rs}_{00}$ and $X^{rs}_{mn}$ are calculated in \cite{GJ1,GJ2,leclair1,leclair2,RSZ4,Feng} and ${\cal N} = 3^{9/2}/2^{6}$. For completeness, it is worth mentioning  that this vertex is also used to make the star product between two string fields

\begin{equation}
| \Phi \star \Psi \rangle_{3} = ~_{1}\langle \Phi| _{2} \langle \Psi| |V_{123} \rangle
\label{star}
\end{equation}

\section{TFD Duplication Rules and The Ground State}

The formalism of Thermal Field Dynamics \cite{ume2,ume1} is a natural extension of general quantum field theories built up in order to describe the thermal effects and statistical properties of a system as an entanglement of its degrees of freedom with a non-interacting identical copy of itself. According to this formalism, one considers the direct product of both Hilbert spaces, and the time evolution is generated by a decoupled hamiltonian operator $\hat{H} \equiv H-\tilde{H}$, where the tilde refers to the copy of system. The operators of the $\widetilde{QFT}$ are constructed from the $QFT$ ones by the \emph{tilde
conjugation rules}, or simply TFD rules \cite{kha5}, defined for all the operators $X, Y, ....$ of the QFT by:

\begin{eqnarray}
( X Y)\widetilde{} &=& \widetilde{X}\widetilde{Y} \nonumber \\
(c\, X + Y) \widetilde{} &=& c^{\ast} \, \widetilde{X}_{} + \widetilde{Y}_{} \nonumber \\
(X^{\dagger })\widetilde{} &=& (\widetilde{X})^{\dagger } \label{tilderules} \\
\lbrack \widetilde{X},Y] &=& 0  \nonumber \\
(\widetilde{X})\widetilde{} &=& \epsilon X \nonumber
\end{eqnarray}
In the last line, $\epsilon=+1(-1)$ for commuting (anti-commuting) fields \cite{ojima}.
This structure is related to a $c^\star$-algebra, and the rules (\ref{tilderules}) may be identified with the modular conjugation of the standard representation \cite{emch}.

In TFD, an extra condition on entangled/thermal ground states is demanded:

\be (i) ~~~~~~~~|\Omega_\theta\rangle \!\rangle = \widetilde{| \Omega_\theta \rangle \!\rangle} \ee where $\theta$ denotes the label on the vacua. Furthermore, all these states are defined to be annihilated by the combination

 \be (ii) ~~~~~~~\widehat{H} \equiv H - \widetilde{H} \ee which is considered to be the generator of time evolution of the duplicated system.
 In particular, using the rules (\ref{tilderules}), one can observe that the canonical open string vacuum
$|\Omega \rangle = c_{1} |0;k \rangle\,$ should be extended a priori as $| \Omega \rangle \!\rangle \equiv | \Omega \rangle \otimes | {\tilde \Omega} \rangle = c_{1} |0;k \rangle \otimes \tilde{c}_{1} |\tilde{0};\tilde{k} \rangle$, and we will see below which specific form this state must have in order to satisfy axioms (i) and (ii).

The last axiom of TFD is the so-called KMS condition that is often expressed as
\begin{eqnarray}
(iii)  ~~~~~~ {\cal O}(x^\mu) | \Omega(\b)\rangle\! \rangle &=& \tilde{\cal O}^\dag(x^\mu\, -\, i \beta^\mu /2)| \Omega(\b)\rangle\! \rangle \nonumber \\
 \langle\!\langle \Omega(\b)| {\cal O}(x^\mu)  &=& \langle\!\langle \Omega(\b)|\tilde{\cal O}^\dag(x^\mu + i \beta^\mu /2) \epsilon \nonumber
\end{eqnarray}
for a spacetime point-dependent operator ${\cal O}$,
where $\beta^\mu$  is a timelike vector and $\b \equiv \sqrt{-\beta^\mu\beta_\mu }$ is the temperature inverse. This is what \emph{defines} the thermal ground state.

In this first paper, we are not going to implement the KMS condition (iii) that breaks the relativistic symmetry and restricts the general mixed states to the equilibrium (thermal) ones. Hence, the resulting theory can be viewed as a generalization of TFD, which can be interpreted as a theory for mixed states and addresses the more general possibility of describing dynamics for states out of thermal equilibrium.

\subsection{The TFD double string vacuum}

As an application of the TFD rules, let us first construct the ground state of the doubled open string.
We start with a physical one-string ground state and another (independent) one for the tilde copy of the state space:
\begin{equation}\label{ambosvacios}
| \Omega \rangle : = c_{1} |0;k \rangle =  e^{ i k \cdot X_0} c_{1} | 0 ; 0 \rangle ~~;~~ | \widetilde{\Omega} \rangle := {\tilde c}_{1} |{\tilde 0};{\tilde k} \rangle = e^{ i {\tilde k} \cdot \widetilde {X_0} } {\tilde c}_{1} | {\tilde 0};  {\tilde 0}\rangle
\end{equation}
 The operator $X_0$ denotes $X(z=\bar{z}=0)$ in the complex plane.
Then the vacuum of the doubled theory shall be defined in general as

\be\label{vacio-general}| \Omega \rangle \!\rangle \equiv | \Omega \rangle \otimes | {\tilde \Omega} \rangle = c_{1} | 0; k \rangle \otimes  {\tilde c}_{1}  | {\tilde 0}; {\tilde k} \rangle\ee
However, notice that not all of these states are vacua of the extended theory; indeed, according to the axiom $(ii)$, the ground states must satisfy the condition $| \Omega \rangle \!\rangle= \widetilde{| \Omega \rangle \! \rangle}$, i.e

\be\label{condition}
c_{1} |0;k \rangle  \otimes {\tilde c}_{1}|{\tilde 0};{\tilde k} \rangle  = - \tilde{c}_{1} \widetilde{\left(|0;k \rangle\right)} \otimes  c_{1} \widetilde{\left(  |{\tilde 0};{\tilde k} \rangle \right)}
\ee

Taking the tilde of both equations in (\ref{ambosvacios})  and using the rules (\ref{tilderules}), we obtain
\begin{equation}\label{v1}
\widetilde{\left(  | \Omega \rangle\right)} : = \tilde{c}_{1} \widetilde{\left(|0;k \rangle\right)} =  e^{ -i k \cdot \widetilde{X}_0} \tilde{c}_{1} \widetilde{| 0 ; 0 \rangle} ~~;~~ \widetilde{\left(| \widetilde{\Omega} \rangle\right)} := - c_{1} \widetilde{\left(  |{\tilde 0};{\tilde k} \rangle \rangle\right)} = - e^{ - i {\tilde k} \cdot X_0 }  c_{1} \widetilde{\left(  | {\tilde 0};  {\tilde 0}\rangle \rangle\right)}
\end{equation}

Then (\ref{condition}) writes as
\begin{equation}
e^{ i k \cdot X_0} c_{1} | 0 ; 0 \rangle \otimes e^{ i {\tilde k} \cdot \widetilde {X_0} } {\tilde c}_{1} | {\tilde 0};  {\tilde 0}\rangle =
 e^{ -i k \cdot \widetilde{X}_0} \tilde{c}_{1} \widetilde{| 0 ; 0 \rangle} \otimes - e^{ -i {\tilde k} \cdot X_0 }  c_{1} \widetilde{\left(  | {\tilde 0};  {\tilde 0}\rangle \rangle\right)}
\end{equation}
that can be rewritten as
\begin{equation}
e^{ i k \cdot X_0} e^{ i {\tilde k} \cdot \widetilde {X_0} } c_{1}{\tilde c}_{1}  | 0 ; 0 \rangle \otimes | {\tilde 0};  {\tilde 0}\rangle =
 e^{ -i k \cdot \widetilde{X}_0} e^{- i {\tilde k} \cdot X_0 } c_{1} \tilde{c}_{1} \widetilde{| 0 ; 0 \rangle} \otimes    \widetilde{\left(  | {\tilde 0};  {\tilde 0}\rangle \rangle\right)}\,
\end{equation}
thus, using that $[X_0 , \widetilde {X_0}]=0 $,
\begin{equation}
e^{ i (k + \tilde{k}) \cdot X_0} e^{ i (k + \tilde{k}) \cdot \widetilde {X_0} } c_{1}{\tilde c}_{1}  | 0 ; 0 \rangle \otimes | {\tilde 0};  {\tilde 0}\rangle = c_{1} \tilde{c}_{1} \widetilde{| 0 ; 0 \rangle} \otimes    \widetilde{\left(  | {\tilde 0};  {\tilde 0}\rangle \rangle\right)}\,,
\end{equation}
which is nothing but
\begin{equation}
 c_{1}{\tilde c}_{1}  | 0 ; k + \tilde{k} \rangle \otimes | {\tilde 0};  k + \tilde{k}\rangle = c_{1} \tilde{c}_{1} \widetilde{| 0 ; 0 \rangle} \otimes    \widetilde{\left(  | {\tilde 0}; {\tilde 0}\rangle \rangle\right)}\, .
\end{equation}
Finally, since the states $\widetilde{| 0 ; 0 \rangle} \,,\,   \widetilde{\left(  | {\tilde 0}; {\tilde 0}\rangle \rangle\right)}$ do not depend on the momenta $k$ or $ \tilde{k}$, we
 we conclude both
\be
\label{k-ktilde}
k + \tilde{k} =0
\ee
and
\be\label{vv2}
\widetilde{| 0 ; 0 \rangle}  =   | {\tilde 0};  {\tilde 0}\rangle\,\,.
\ee

The conclusion, then, is that the ground state of the (TFD) doubled open string theory (\ref{vacio-general}) reads
\be\label{doblevacuo-final}
 | \Omega \rangle \!\rangle \equiv | \Omega \rangle \otimes | {\tilde \Omega} \rangle = c_{1} | 0; k \rangle \otimes  {\tilde c}_{1}  | {\tilde 0}; - k \rangle \, ,
 \ee
and, furthermore, by virtue of (\ref{ambosvacios}), (\ref{v1}) and (\ref{vv2}), we have that  $\widetilde{| \Omega \rangle} = | {\tilde \Omega} \rangle$.

Therefore, the axiom $(ii)$ is automatically satisfied by the string Hamiltonian operator $L_0$. In fact, $\hat{L}_0 \equiv L_0 - \widetilde{L}_0 $ annihilates the state (\ref{doblevacuo-final}) since $ |\tilde{k}|^2=|-k|^2 =|k|^2$.

\section{Extending OSFT}

The first ingredient of TFD is the duplication of the space of states of a quantum theory. This allows us to describe all the states of the system, including density matrices, as pure states. The effects of the statistical mixing are encoded in the entanglement between both parts of the extended theory.

Based on this approach, we are going to duplicate the string Fock space towards a future formulation of a finite temperature string field theory. Our string field will then be constructed as an expansion on a doubled Fock space. The usual string field is an element of an algebra ${\cal A}$, which can be interpreted as a set of (wave) functionals of a string configuration in space-time $\Phi[X(\sigma)]$. Its extension will be a space ${\cal A} \otimes \tilde{{\cal A}}$, whose elements can now be described as functionals of \emph{two} string configurations $\widehat{\Phi}[X(\sigma), \widetilde{X}(\widetilde{\sigma})]$. We will interpret this new object as encoding the information on general (pure or mixed) string states, often described by density matrices (see eq. (\ref{tr1})).

This viewpoint constitutes a radical difference with regard to previous TFD formulations \cite {leblanc,leb1} where, the string field is described as its decomposition in terms of conventional pointwise fields, and thus, the TFD rules simply duplicate them. Here, the duplication is viewed on the string field itself, and one might expect different consequences on the field spectrum.

As just said, the extension of the string field is immediate by considering the tensor product of both basis. Having into account (\ref{k-ktilde}), we can represent this as:

\begin{equation}
| \Phi \rangle \!\rangle = \int \frac{d^{26}k}{(2 \pi)^{26}} \left[ t(k) + A_{\mu}(k) \alpha^{\mu}_{-1} + B_{\mu}(k) {\tilde \alpha}^{\mu}_{-1} +  C_{\mu \nu}(k) \alpha^{\mu}_{-1} {\tilde \alpha}^{\nu}_{-1} + \cdots \right] | \Omega \rangle \!\rangle
\label{entosftstate}
\end{equation}

According to conventional TFD, one shall also consider a tilde copy of this string field constructed from it by the tilde conjugation rules:
\begin{eqnarray}
| \widetilde{\Phi} \rangle \!\rangle = \int \frac{d^{26}k}{(2 \pi)^{26}}  \left[ \tilde{t}(k) + \widetilde{A}_{\mu}(k) {\tilde \alpha}^{\mu}_{-1} + \widetilde{B}_{\mu}(k) \alpha^{\mu}_{-1} + \widetilde{C}_{\mu \nu}(k) {\tilde \alpha}^{\mu}_{-1} \alpha^{\nu}_{-1} \cdots \right] | \Omega \rangle \!\rangle
\label{entosftstate-til}
\end{eqnarray}
where we have implicitly assumed that the component fields shall be canonically quantized afterwards,
 so the fields $ t, A_{\mu}, C_{\mu \nu}, \dots$ must be considered independent from their tilde partners.

Following the TFD construction, the theory is defined by the difference between two (non-interacting) string field theories

\begin{equation}
\hat{S}\,[\Phi, \widetilde{\Phi}] = S\,[\Phi] - \widetilde{S}\,[\widetilde{\Phi}],
\label{TFDaction}
\end{equation}
where the first term is the usual (open) string field theory action properly extended to the space of configurations ${\cal A} \otimes \tilde{{\cal A}}$  \footnote{The SFT action must be extended to be a well defined functional of fields (\ref{entosftstate}).}. And the second one, is derived from this by using the tilde conjugation rules.

Formally, this is

\begin{equation}
{\hat S} =  \frac{1}{2} \langle\! \langle \Phi | Q_{B} | \Phi \rangle \!\rangle_{ext} - \frac{1}{2}  \langle\! \langle {\widetilde \Phi} | {\widetilde Q}_{B} | {\widetilde \Phi}  \rangle \!\rangle_{ext} + \frac{g}{3}\langle\!\langle \Phi, \Phi \star \Phi \rangle \!\rangle_{ext} - \frac{g}{3} \langle\!\langle \widetilde{\Phi}, \widetilde{\Phi} \star \tilde{\Phi} \rangle \!\rangle_{ext}
\label{formal-ext-action}
\end{equation}

From now on, we omit \emph{ext} to denote the extension of the scalar product to the space ${\cal A} \otimes \tilde{{\cal A}}$, and we simply denote it by the double bracket.
Although (\ref{formal-ext-action}) describes the correct structure of the extended action,  computationally, one shall give a prescription to extend the scalar product to act on $\widehat{{\cal A}}\equiv {\cal A} \otimes \tilde{{\cal A}}$, and its tensor products $\otimes_i \widehat{{\cal A}}_i$, so as to extend the product $\star$ on \emph{two} doubled string spaces $\widehat{{\cal A}}_1\otimes\widehat{{\cal A}}_2$. We take the reflector state for this space simply as $\langle \!\langle R_{12}| \equiv \langle R_{12}|\otimes\langle \widetilde{R}_{12}|$, where $\langle \widetilde{R}_{12}| \equiv \langle R_{\tilde{1}\tilde{2}}|$,
 and, using the rules above, it can be verified that
 \be
 \widetilde{\langle \!\langle R_{12}|} =  \langle \!\langle R_{12}| \,.
 \ee

By minimizing (\ref{formal-ext-action}) with respect to the fields $\Phi$ and $\widetilde{\Phi}$, we obtain two decoupled equations of motion
\begin{equation}
 Q_{B}| \Phi \rangle \!\rangle = g |\Phi \, \star\,\Phi \rangle \!\rangle
\label{eqmotion}
\end{equation}
and
\begin{equation}
 \widetilde{Q}_{B} | \widetilde{\Phi} \rangle \!\rangle = g |\widetilde{\Phi} \, \star\,\widetilde{\Phi} \rangle \!\rangle
\label{eqmotion-til}
\end{equation}
whose classical solutions would describe the states of the theory. From now on, we will focus on the \emph{free} OSFT, which describes the asymptotic states or the weak coupling limit of the above theory, so the r.h.s. of these two equations should be interpreted only as formal expressions.
 In the last section, we shall discuss about the possible form of the interaction term.

\subsection{The Free Theory}

Let us write the kinetic term of the extended free OSFT in the following way
\begin{equation}
S[\Phi] =  \frac{1}{2} \langle \! \langle \Phi | Q_{B} | \Phi \rangle \!\rangle\, ,
\label{kinetic-term}
\end{equation}
so the full TFD action is
\begin{equation}
{\hat S} =  S[\Phi] - \tilde{S}[\widetilde{\Phi}]\,.
\label{freeextaction}
\end{equation}

We must then give a prescription for the internal product $\langle\! \langle \Phi_1 | \Psi_2 \rangle \!\rangle $ on $\widehat{{\cal A}} $, in order to have the correct ghost number in the kinetic term. For instance, let us consider some generic operator ${\cal O}$ acting on one copy of the usual space ${\cal A}$, then the Riemann-Roch theorem implies that the product $\langle \Phi_1 | {\cal O}\Psi_2 \rangle $ is nontrivial iff the ghost number in this product is 3, since it is to be evaluated on the disk. So, for instance, if $|\Phi_1\rangle$ and $|\Psi_2\rangle $ are ordinary open string fields $gh (|\Phi\rangle)  = gh (|\Psi\rangle)  = 1$, then one shall have $gh ({\cal O}) = 1 $. Now, let us consider an operator $\widehat{{\cal O}}$ on the doubled open string space $\widehat{{\cal A}}$. The ghost numbers are $gh (|\Phi\rangle \!\rangle) = \widetilde{gh} (|\Phi\rangle \!\rangle) = gh (|\Psi\rangle \!\rangle) = \widetilde{gh} (|\Psi\rangle \!\rangle) = 1$, and the cited theorem implies that the product  $ \langle\! \langle \Phi | \widehat{{\cal O}}\,\Psi \rangle \!\rangle$ is nontrivial only if $gh (\widehat{{\cal O}}) = \widetilde{gh} (\widehat{{\cal O}}) =  1$.

This is essentially the situation of defining the kinetic term (\ref{kinetic-term}), $\langle \!\langle \Phi | Q^{}_{B} | \Phi \rangle \!\rangle\,$,
 where $ gh  (Q^{}_{B}) = 1 $.
The extended \emph{reflection} map
$\langle\!\langle R_{12}| : \widehat{{\cal A}}_1 \to \widehat{{\cal A}}^*_2  $ can be thought of as a metric
$\langle\!\langle R_{12}| : \widehat{{\cal A}}_1\otimes\widehat{{\cal A}}_2 \to \Re $ defining the internal product.
 The kinetic term (\ref{kinetic-term}) would then be expressed as

 \be \left(\langle \!\langle R_{12}| \Phi_{(1)}\rangle \!\rangle \right) | Q^{(2)}_{B} | \Phi_{(2)} \rangle \!\rangle \ee
 where the indices $1, 2, \dots$ stand for different (doubled) string space copies $\widehat{{\cal A}}_{1, 2, \dots}$. Notice that the ghost number is not saturated unless we insert some operator ${\cal G}^{(2)}$ such that the unique non-vanishing ghost number is $\widetilde{gh}_2 {\cal G} = 1 $. An obvious candidate to this is, of course, the operator $\tilde{c}^{(2)}_0$.

 Therefore, we see that the correct formula to define this is \footnote{A study on how this insertion should be, for general ghost numbers, will be presented in a forthcoming work.}

  \be S[\Phi]\equiv \frac{1}{2} \left(\langle \!\langle R_{12}| | \Phi_{(1)}\rangle \!\rangle \right)  {\cal G}^{(2)} Q^{(2)}_{B} | \Phi_{(2)} \rangle \!\rangle \ee

So our proposal is to extend the internal product to the doubled Hilbert space by means of
\be
\langle \! \langle \Phi_1 | \Psi_2 \rangle \!\rangle \equiv \langle \! \langle R_{12} | |\Phi_1 \rangle \!\rangle {\cal G}^{(2)} | \Psi_2 \rangle \!\rangle
\label{innerprod}
\ee
where $\langle \! \langle \Phi_1 |$ has $gh = \widetilde{gh} =1$, whereas the object $| \Psi_{2} \rangle \! \rangle$ has $gh = 2$ and $\widetilde{gh} = 1$. In order to capture a more general prescription, but also for simplicity, we take this to be a generic linear combination of ghost operators
\be\label{innerprodinsert}
{\cal G} = l\, c_0 + \tilde{l} \, \tilde{c}_0 \,, ~~~~~l, \tilde{l} \in \Re .
\ee
Notice that all internal products of this family, parameterized by the real numbers $l, \tilde{l}$, are indeed \emph{non-degenerate}.

Within this prescription, the action (\ref{kinetic-term}) reads
\begin{equation}
S[\Phi] =  \frac{1}{2} \left(\langle \! \langle R_{12} | \Phi\rangle \!\rangle\right) {\cal G} \,Q_{B} | \Phi \rangle \!\rangle\, .
\label{kinetic-term-R}
\end{equation}
Then, using $R_{12}=  \widetilde{R}_{12}$, and the TFD rules, the tilde action is simply
\begin{equation}
\tilde{S}[\tilde{\Phi}] =   \frac{1}{2} \left(\langle \!\langle R_{12} |  \tilde{\Phi}\rangle \!\rangle\right) \widetilde{{\cal G}} \,\tilde{Q}_{B} | \tilde{\Phi} \rangle \!\rangle\, ,
\label{kinetic-term-R-tilde}
\end{equation}
where interestingly, the tilde action involves a different (tilde) internal product.
 Therefore, the equations of motion derived by variating the action $\hat{S}[\Phi,{\widetilde \Phi}]$ with respect to the kets $| \Phi\rangle \!\rangle$ and $| \tilde{\Phi}\rangle \!\rangle$ respectively are

\begin{equation}
Q_{B} | \Phi \rangle \!\rangle = 0 ~~~ \& ~~~ \widetilde{Q}_{B} | {\widetilde \Phi} \rangle \!\rangle = 0
\label{extEOM}
\end{equation}
Let us remark that these equations of motion are the same and are independent on the choice of the internal product. However, let us end this section by mentioning that there is a particularly symmetric choice of the operator ${\cal G}_c \equiv l \hat{c}_0 = l ( c_0 - \tilde{c}_0 ) $, which is in line with the TFD-extension of the operators. Then $ \widetilde{{\cal G}}_c = - {\cal G}_c $ is verified, and so the tilde corresponding product is the same with inverted signal. This choice is referred to as the \emph{canonical product}. So the extended OSFT defined with the canonical product results
\begin{equation}
{\hat S}_{canonical} =  \frac{l}{2} \left( \langle\!\langle R_{12} | \Phi \rangle\!\rangle \right) \hat{c}_0 Q_{B} | \Phi \rangle \!\rangle\, + \, \frac{l}{2} \left( \langle\!\langle R_{12} |  \tilde{\Phi}\rangle \!\rangle \right)  \hat{c}_0 \tilde{Q}_{B} | \tilde{\Phi} \rangle \!\rangle\,\,.
\label{canonicalaction}
\end{equation}
Later, we will see that this encodes an interesting property.

\section{Ground States}

 In this section, we will study the theory that describes the \emph{generally entangled ground states} of the doubled-OSFT (or simply  $\Omega$-states). By tracing out the tilde copy degrees of freedom, these (mixed) states can be equivalently described by density matrices encoding the statistical properties of the open string field. In the present approach, one can restrict the theory to these states by only imposing the  axioms (i, ii) of TFD. Then, in a following step, by imposing the KMS condition (axiom (iii)), one would obtain the specific OSFT thermal vacuum. Thus, here we will investigate the space of $\Omega$-states, which satisfy the axioms (i,ii).

As seen in section 3, the extended canonical string vacuum is
$| \Omega \rangle \!\rangle = | \Omega, {\tilde \Omega} \rangle \!\rangle =| \Omega \rangle \otimes | {\tilde \Omega} \rangle = c_{1} | 0; k \rangle \otimes  {\tilde c}_{1}  | {\tilde 0}; -k \rangle$
 and it satisfies both
\be
(L_0 - 1)\, |\Omega \rangle \!\rangle = 0~~,~(\tilde{L}_0 - 1)\, |\Omega\rangle \!\rangle=0\,\,.
\ee
It then automatically satisfies $L_0 - \tilde{L}_0 \, | \Omega \rangle \!\rangle = 0$.
Since $\hat{L}_0 \equiv L_0 - \tilde{L}_0$ is the total Hamiltonian in a TFD formulation, one then demands that this property characterize the most general entangled ground states (axiom (ii)):
\be\label{axiomii}
\;\;\;\;\hat{L}_0  \, | \Omega(\theta) \rangle \!\rangle = L_0 - \tilde{L}_0 \, | \Omega(\theta) \rangle \!\rangle = 0\,.
\ee
Because there is an infinite-dimensional space of solutions for this equation, we have labeled these by parameters $\theta$'s which will be interpreted below.

In addition, as required by the TFD axiom (i), we also demand the invariance
\begin{equation}\label{axiomi}
 \;\;\;\;| \Omega(\theta)\rangle \!\rangle = \widetilde{|\Omega (\theta) \rangle \!\rangle}\,.
\end{equation}
These two axioms are what we could minimally require to define generally entangled ground states of the extended (free) OSFT
 (\ref{freeextaction})
 \begin{equation}
{\hat S} =  \frac{1}{2} \langle \!\langle \Phi | Q_{B} | \Phi \rangle \!\rangle - \frac{1}{2}  \langle \!\langle {\widetilde \Phi} | {\widetilde Q}_{B}| {\widetilde \Phi}  \rangle \!\rangle \, .
\end{equation}
The equations of motion (\ref{extEOM})
\begin{equation}
Q_{B} | \Phi \rangle \!\rangle = 0 ~~~ \& ~~~ \widetilde{Q}_{B} | {\widetilde \Phi} \rangle \!\rangle = 0\,
\end{equation}
are satisfied for all string fields $| \Phi \rangle \!\rangle$, $| \widetilde{\Phi} \rangle \!\rangle$. In particular they hold for ground states defined according to (\ref{axiomii}) and (\ref{axiomi}). Then, using (\ref{axiomi}), we obtain:
\begin{equation}
Q_{B} \, \pm\, \widetilde{Q}_{B} \,\,| \Omega(\theta) \rangle \!\rangle = 0
\label{Qpm}
\end{equation}
which, on the other hand, can be derived from another equivalent free (effective) action for $\Omega$-fields:
\begin{equation}\label{csft-action}
{\cal S}_\pm[\Omega(\theta)] =  \frac{1}{2} \left( \langle \!\langle R_{12}| |\Omega(\theta) \rangle \! \rangle \right)  {\cal G}_\pm \, Q_\pm  |  \Omega(\theta)  \rangle \!\rangle
\end{equation}
where $Q_\pm \equiv Q_{B} \, \pm\, \widetilde{Q}_{B}$ and ${\cal G}_\pm \equiv   \widetilde{{\cal G}} \, \pm\, {\cal G}$.

Therefore, by virtue of (i) and (ii), these fields can be written at leading energy level as
\begin{eqnarray}\label{vacua-theta-fields}
| \Omega(\theta)\rangle\! \rangle = \int \frac{d^{26}k}{(2 \pi)^{26}}  \left[ t(k) +
    C_{\mu \nu}(k) \alpha^{\mu}_{-1} {\tilde \alpha}^{\nu}_{-1} + \cdots \right] | \Omega \rangle \!\rangle \, ;
\end{eqnarray}
with $t(k) = \tilde{t}(k)$ and $C_{\mu \nu} (k) = \tilde{C}_{\nu \mu}(k)$. The vacua can thus be characterized by $ (t, C_{\mu \nu}, \dots)$ and, hence, this collection of fields can be identified with the parameters $\theta$'s themselves.

Notice that these fields associated with the $\Omega$-states appear without their tilde partners. In other words, the TFD duplication will only produce a tilde correspondent of the ordinary open string fields, but not for the background fields such as $C_{\mu \nu}$. Therefore, if we decompose this field in its irreducible components $C_{\mu \nu} = g_{\mu \nu} + B_{\mu \nu} + \phi \eta_{\mu \nu}$, we can conjecture that the field $g_{\mu \nu}$ precisely describes gravitons, and that thermal (open) string fields might describe the gravitational field, and more hopefully, closed strings.

The main result of this section is the observation that, upon the appropriate identifications of the string degrees of freedom, this effective theory can be \emph{identified} with certain formulation of (free)  Closed String Field Theory (CSFT) \cite{Zwiebach}.

In fact,  if we identify the holomorphic/anti-holomorphic part (a.h.) of the closed string degrees of freedom with the tilde/non-tilde open strings respectively, one obtains that both theories are coincident.

In particular, the extended reflector $\langle\!\langle R_{12}|\equiv\langle R_{12}|\langle R_{\tilde{1}\tilde{2}}|$ results to be suggestively similar to the CSFT one \cite{Zwiebach}:

\begin{eqnarray}
\langle R^c_{12}| &=& \int \frac{d^{26}k}{(2\pi)^{26}} (_{1} \langle 0, \tilde{0} ;k| c_{-1}\bar{c}_{-1} \otimes ~_{2} \langle 0, \tilde{0};-k|  c_{-1}\bar{c}_{-1}) (c_{0}^{-(1)} + c_{0}^{-(2)})(c_{0}^{+(1)} + c_{0}^{+(2)}) \nonumber \\
&\times& {\rm exp} \left( \sum_{n=1}^{\infty}  \left[ a_{n}^{(1)} a_{n}^{(2)} + c_{n}^{(1)} b_{n}^{(2)} +
 c_{n}^{(2)} b_{n}^{(1)} + a.h. \right] \right)
\label{reflectorC}
\end{eqnarray}
with a.h. denoting the antiholomorphic part and where
\begin{eqnarray}
 c_{0}^{\pm} = (c_{0} \pm \bar{c}_{0}) /2\,.
\label{identif-C-O}
\end{eqnarray}
We see that both definitions of the reflector coincide \emph{up to a twist} on the string ``(2)'' given by:
\begin{eqnarray}
\left(  a_{n}^{(2)} , b_{n}^{(2)} , c_{n}^{(2)} \right) \rightarrow \left( (-1)^{n} a_{n}^{(2)} , (-1)^{n} b_{n}^{(2)} , (-1)^{n} c_{n}^{(2)} \right)
\label{twist}
\end{eqnarray}
which is a canonical transformation.

Furthermore, the equation of motion (\ref{Qpm}), with $Q_+$, coincides with that of (free) CSFT \cite{Zwiebach}, and (\ref{axiomii}) can be identified with the level matching condition, so that both theories are (on-shell) equivalent.
  Moreover, such equivalence becomes \emph{off-shell} if, and only if, the inner product (\ref{innerprod}) is the \emph{canonical} one, i.e, the TFD-extended action is (\ref{canonicalaction}).
  In fact, for states satisfying the constraint (ii), eq. (\ref{canonicalaction}) expresses as
\begin{equation}
{\hat S}_{canonical}[\Omega] =   \frac{1}{2} \left( \langle \!\langle R_{12} | | \Omega\rangle \!\rangle\right) \hat{c}_0 (Q_{B} + \tilde{Q}_{B}) | \Omega \rangle \!\rangle\,\,.
\label{canonical-CSFT}
\end{equation}
which is nothing but the free-CSFT action.

  So, the result found here is that the $\Omega$-states can be identified with the asymptotic states of (free) CSFT.
In the following section, we are going to study the field spectrum of double-OSFT and explicitly verify that.

\section{Spectrum Analysis of the Free Theory}

In the first quantized theory of strings, one way to get the spectrum of the free theory is to use the BRST cohomology so that physical states are those in the cohomology of the $Q_{B}$ operator. Here, we will pursue the same path.

The open string BRST operator is

\begin{equation}
Q_{B} = \sum_{n=-\infty}^{\infty} c_{n}L_{-n}^{(M)} + \sum_{m,n=-\infty}^{\infty} \frac{(m-n)}{2} :c_{m}c_{n}b_{-m-n}: - c_{0}
\label{OpenBRST}
\end{equation}
where $M$ stands for the matter part and $: \cdots :$ means normal ordering with respect to the $|\Omega\rangle$ vacuum. The TFD rules give us then

\begin{equation}
\widetilde{Q}_{B} = \sum_{n=-\infty}^{\infty} {\tilde c}_{n} {\tilde L}_{-n}^{(M)} + \sum_{m,n=-\infty}^{\infty} \frac{(m-n)}{2} :{\tilde c}_{m} {\tilde c}_{n} {\tilde b}_{-m-n}: - {\tilde c}_{0}
\label{OpenBRSTtilde}
\end{equation}

Let us then analyze what happens when we apply the extended BRST charge to the doubled field. For that, we define the level of the field by the pair of numbers $(N,{\tilde N})$, where $N$ is the eigenvalue of the basis state with respect to the number operator present in $L_{0}$, and analogously for the tilde operator. We will do the analysis for the first few levels.

\subsection{Cohomology of $Q_{B}$ and ${\tilde Q}_{B}$}

In SFT, we build the string field as a linear combination of the physical states of the first quantised string. These are obtained by various ways, one of them is by the cohomology of the BRST operator. In this section, we are going to study the cohomology of the extended BRST operators and see what are the physical states.

Let us start by the cohomology of the $Q_{B}$ operator:

\subsection*{\underline{Level (0,0)}}

In this level, a general state is written as

\begin{equation}
| \Phi^{(0,0)} \rangle \!\rangle = t(k) | \Omega \rangle \! \rangle
\end{equation}
so that

\begin{equation}
Q_{B}  | \Phi^{(0,0)} \rangle \!\rangle =  t(k)  Q_{B} | \Omega \rangle \! \rangle = t(k) (\alpha' k^{2} - 1) c_{0} | \Omega \rangle \! \rangle
\end{equation}
For this state to be closed, it is clear that $k^{2} = 1/\alpha'$. It can also be seen that this state cannot be written as the BRST operator acting on another state. Hence, it is never exact. The conclusion is then that the physical state at this level is a tachyon state.

\subsection*{\underline{Level (1,0)}}

At this level, the general state is

\begin{equation}
| \Phi^{(1,0)} \rangle \!\rangle = \left[ A_{\mu}(k) \alpha^{\mu}_{-1} + \beta(k) c_{-1} + \gamma(k) b_{-1} \right] | \Omega \rangle \! \rangle
\end{equation}
Acting with the $Q_{B}$ operator, we get

\begin{eqnarray}
Q_{B} | \Phi^{(1,0)} \rangle \!\rangle &=& \left[ \sqrt{\frac{\alpha'}{2}} k^{\mu} A_{\mu}(k)c_{-1} + (\alpha' k^{2}) A_{\mu}(k) \alpha^{\mu}_{-1}c_{0} + \sqrt{\frac{\alpha'}{2}} \gamma(k) k_{\mu} \alpha^{\mu}_{-1} \right. \nonumber \\
 &+& \left. (\alpha' k^{2}) \gamma(k) c_{0}b_{-1} - (\alpha' k^{2}) \beta (k) c_{-1}c_{0} \right] | \Omega \rangle \! \rangle
\end{eqnarray}
This state is closed if $k^{2} = k^{\mu} A_{\mu} = \gamma = 0$ and $\beta$ is free. An exact state should satisfy $| \Psi^{(1,0)} \rangle = Q_{B} |\Phi^{(1,0)} \rangle$. We get then

\begin{eqnarray}
| \Psi^{(1,0)} \rangle \!\rangle &=& \left[ A'_{\mu}(k) \alpha^{\mu}_{-1} + \beta'(k) c_{-1} + \gamma' (k) b_{-1} \right] | \Omega \rangle \! \rangle = Q_{B} | \Phi^{(1,0)} \rangle \!\rangle =\nonumber \\
&=& \left[ \sqrt{\frac{\alpha'}{2}} k^{\mu} A_{\mu}(k)c_{-1} + (\alpha' k^{2}) A_{\mu}(k) \alpha^{\mu}_{-1}c_{0} + \sqrt{\frac{\alpha'}{2}} \gamma(k) k_{\mu} \alpha^{\mu}_{-1} \right. \nonumber \\
&+& \left. (\alpha' k^{2}) \gamma(k) c_{0}b_{-1} - (\alpha' k^{2}) \beta (k) c_{-1}c_{0} \right] | \Omega \rangle \! \rangle
\end{eqnarray}
Since $k^{\mu} A_{\mu} \neq 0$, otherwise $|\Phi^{(1,0)} \rangle$ would be closed, we get that for a state to be exact we must then have $k^{2} = 0$, $\gamma' = 0$ and

\begin{equation}
A'_{\mu} (k) = \sqrt{\frac{\alpha'}{2}} \gamma (k) k_{\mu} ~~;~~ \beta' (k) = \sqrt{\frac{\alpha'}{2}} k^{\mu} A_{\mu} (k)
\end{equation}
It means that for a state to be closed and not exact we should have $k^{2} = 0$, $\gamma = 0$, $k^{\mu} A_{\mu} = 0$ and $\beta \neq \sqrt{\alpha' /2} (k^{\mu} A_{\mu}) = 0$. So $c_{-1} | \Omega \rangle \! \rangle$ is an exact state for it can be written as

\begin{equation}
c_{-1} | \Omega \rangle \! \rangle = \frac{1}{\beta} Q_{B} \left( A_{\mu} \alpha^{\mu}_{-1} | \Omega \rangle \! \rangle \right)
\end{equation}
We also get a gauge invariance $A'_{\mu} \simeq A_{\mu} + \sqrt{\alpha' /2} ~\gamma k_{\mu}$. The physical state at this level is then

\begin{equation}
| \Phi^{(1,0)} \rangle \!\rangle = A_{\mu}(k,{\tilde k}) \alpha^{\mu}_{-1} | \Omega \rangle \! \rangle
\end{equation}
with $k^{2} = 0$ and the aforementioned gauge invariance, that is, a $U(1)$ gauge field.

\subsection*{\underline{Level (0,1)}}

Now we have

\begin{equation}
| \Phi^{(0,1)} \rangle \!\rangle = \left[ B_{\mu}(k) {\tilde \alpha}^{\mu}_{-1} + \zeta(k) {\tilde c}_{-1} + \xi (k) {\tilde b}_{-1} \right] | \Omega \rangle \! \rangle
\end{equation}
so that

\begin{equation}
Q_{B} | \Phi^{(0,1)} \rangle \!\rangle = (\alpha' k^{2} - 1) \left[ B_{\mu}(k) {\tilde \alpha}^{\mu}_{-1} + \zeta(k) {\tilde c}_{-1} +  \xi (k) {\tilde b}_{-1} \right] c_{0} | \Omega \rangle \! \rangle
\end{equation}
The state is closed only if $k^{2} = 1/ \alpha'$. If $k^{2} \neq 1/ \alpha'$, we see that

\begin{equation}
| \Phi^{(0,1)} \rangle \!\rangle = \frac{1}{\alpha' k^{2} -1} Q_{B} | \Phi^{(0,1)} \rangle \!\rangle
\end{equation}
that is, it is an exact state. Hence, every state in this level is physical as long as $k^{2} = 1/ \alpha'$.

\subsection*{\underline{Level (1,1)}}

The general state is now

\begin{eqnarray}
| \Phi^{(1,1)} \rangle \!\rangle &=& \left[ C_{\mu \nu}(k) \alpha^{\mu}_{-1}{\tilde \alpha}^{\nu}_{-1} + v_{\mu} (k) \alpha^{\mu}_{-1} {\tilde b}_{-1} + w_{\mu} (k) \alpha^{\mu}_{-1} {\tilde c}_{-1} + u_{\nu} (k) b_{-1} {\tilde \alpha}^{\nu}_{-1} \right. \nonumber \\
&+&   z_{\nu} (k) c_{-1} {\tilde \alpha}_{-1}^{\nu} + D (k) b_{-1} {\tilde b}_{-1} + F (k) c_{-1} {\tilde b}_{-1} + H (k) b_{-1} {\tilde c}_{-1} \nonumber \\
&+& \left. M (k) c_{-1} {\tilde c}_{-1} \right] | \Omega \rangle \! \rangle
\label{Phi11}
\end{eqnarray}
Now

\begin{eqnarray}
Q_{B} | \Phi^{(1,1)} \rangle \! \rangle &=& \left[ (\alpha' k^{2}) C_{\mu \nu} \alpha^{\mu}_{-1} {\tilde \alpha}^{\nu}_{-1} c_{0} + \sqrt{\frac{\alpha'}{2}} k^{\mu} C_{\mu \nu} c_{-1} {\tilde \alpha}^{\nu}_{-1} + (\alpha' k^{2}) v_{\mu} \alpha^{\mu}_{-1} c_{0} {\tilde b}_{-1} \right. \nonumber \\
&+& \sqrt{\frac{\alpha'}{2}} k^{\mu} v_{\mu} c_{-1} {\tilde b}_{-1} + (\alpha' k^{2}) w_{\mu} \alpha^{\mu}_{-1} c_{0} {\tilde c}_{-1} + \sqrt{\frac{\alpha'}{2}} k^{\mu} w_{\mu} c_{-1} {\tilde c}_{-1} \nonumber \\
&+& \sqrt{\frac{\alpha'}{2}} k_{\mu} u_{\nu} \alpha^{\mu}_{-1} {\tilde \alpha}^{\nu}_{-1} + (\alpha' k^{2})  u_{\nu} {\tilde \alpha}^{\nu}_{-1} c_{0} b_{-1} + \sqrt{\frac{\alpha'}{2}} D k_{\mu} \alpha^{\mu}_{-1} {\tilde b}_{-1} \label{QPhi11} \\
&+& (\alpha' k^{2}) D c_{0} b_{-1} {\tilde b}_{-1} + \sqrt{\frac{\alpha'}{2}} H k_{\mu} \alpha^{\mu}_{-1} {\tilde c}_{-1} + (\alpha' k^{2}) H c_{0} b_{-1} {\tilde c}_{-1} \nonumber \\
&-& \left. (\alpha' k^{2})  z_{\nu} {\tilde \alpha}^{\nu}_{-1} c_{-1} c_{0} - (\alpha' k^{2})  F c_{-1} c_{0} {\tilde b}_{-1} - (\alpha' k^{2}) M c_{-1} c_{0} {\tilde c}_{-1} \right] | \Omega \rangle \! \rangle \nonumber
\end{eqnarray}
For this state to be closed, we should have

\begin{equation}
k^{2} = 0 ~;~ k^{\mu} C_{\mu \nu} = 0 ~;~ k^{\mu} v_{\mu} = 0 ~;~ k^{\mu} w_{\mu} = 0 ~;~
u_{\nu} = 0 ~;~ D =0 ~;~ H =0
\end{equation}
and $z_{\nu}$, $F$ and $M$ all free.

Now we should look at the exact states. Using the same procedure we used above and comparing (\ref{Phi11}) and (\ref{QPhi11}), we see that an exact state should satisfy $k^{2} = 0$, $u'_{\nu} = D' = H' = 0$ and

\begin{eqnarray}
&&C'_{\mu \nu} = \sqrt{\frac{\alpha'}{2}} k_{\mu} u_{\nu} ~;~ v'_{\mu} = \sqrt{\frac{\alpha'}{2}} D k_{\mu} ~;~ w'_{\mu} = \sqrt{\frac{\alpha'}{2}} H k_{\mu} \\
 && z'_{\nu} = \sqrt{\frac{\alpha'}{2}}  k^{\mu} C_{\mu \nu} ~;~ F' = \sqrt{\frac{\alpha'}{2}} k^{\mu} v_{\mu} ~;~ M' = \sqrt{\frac{\alpha'}{2}} k^{\mu} w_{\mu} \nonumber
\end{eqnarray}
To get a state that is closed but not exact, we should then have

\begin{eqnarray}
k^{2} = 0 ~;~ k^{\mu} C_{\mu \nu} = 0 &;& k^{\mu} v_{\mu} =0 ~;~ k^{\mu} w_{\mu} = 0 \nonumber \\
u_{\nu} = 0 ~;~ D = 0 &;& H = 0 ~;~ z_{\nu} \neq \sqrt{\frac{\alpha'}{2}} k^{\mu} C_{\mu \nu} = 0 \\
F \neq \sqrt{\frac{\alpha'}{2}} k^{\mu} v_{\mu} = 0 &;& M \neq \sqrt{\frac{\alpha'}{2}} k^{\mu} w_{\mu} = 0 \nonumber
\end{eqnarray}
We come to the conclusion that the physical state at this level is

\begin{equation}
| \Phi^{(1,1)} \rangle \! \rangle = \left[ C_{\mu \nu}(k) \alpha^{\mu}_{-1}{\tilde \alpha}^{\nu}_{-1} + v_{\mu} (k) \alpha^{\mu}_{-1} {\tilde b}_{-1} + w_{\mu} (k) \alpha^{\mu}_{-1} {\tilde c}_{-1}  \right] | \Omega \rangle \! \rangle
\end{equation}
with the following gauge invariances

\begin{eqnarray}
&& C'_{\mu \nu} \simeq C_{\mu \nu} + \sqrt{\frac{\alpha'}{2}} k_{\mu} u_{\nu} \nonumber \\
&& v'_{\mu} \simeq v_{\mu} + \sqrt{\frac{\alpha'}{2}} D k_{\mu} \\
&& w'_{\mu} \simeq w_{\mu} + \sqrt{\frac{\alpha'}{2}} H k_{\mu} \nonumber
\end{eqnarray}
We can decompose the tensor $C_{\mu \nu}$ in its irreducible parts so that $C_{\mu \nu} = g_{\mu \nu} + B_{\mu \nu} + \phi \eta_{\mu \nu}$, that is the direct sum of its symmetric, antisymmetric and traceless parts. We see then that the gauge invariances become

\begin{eqnarray}
&& g'_{\mu \nu} \simeq g_{\mu \nu} + \sqrt{\frac{\alpha'}{2}} (k_{\mu} u_{\nu} + k_{\nu} u_{\mu}) \nonumber \\
&& B'_{\mu \nu} \simeq B_{\mu \nu} + \sqrt{\frac{\alpha'}{2}} (k_{\mu} u_{\nu} - k_{\nu} u_{\mu}) \nonumber \\
&& \phi' \simeq \phi + \sqrt{\frac{\alpha'}{2}} k^{\mu} u_{\mu} \\
&& v'_{\mu} \simeq v_{\mu} + \sqrt{\frac{\alpha'}{2}} D k_{\mu} \nonumber \\
&& w'_{\mu} \simeq w_{\mu} + \sqrt{\frac{\alpha'}{2}} H k_{\mu} \nonumber
\end{eqnarray}
As one can see, we obtain at this level two $U(1)$ gauge fields and what it seems to be the fields of the graviton, the Kalb-Ramond field and the dilaton with their respective gauge invariances.

Hence, up to level (1,1), the extended string field is

\begin{eqnarray}
| \Phi \rangle \!\rangle &=& \int \frac{d^{26}k}{(2 \pi)^{26}} \left[ t(k) + A_{\mu}(k) \alpha^{\mu}_{-1} + B_{\mu}(k) {\tilde \alpha}^{\mu}_{-1} + \zeta(k) {\tilde c}_{-1} + \xi (k) {\tilde b}_{-1} \right. \nonumber \\
 &+& \left. C_{\mu \nu}(k) \alpha^{\mu}_{-1}{\tilde \alpha}^{\nu}_{-1} + v_{\mu} (k) \alpha^{\mu}_{-1} {\tilde b}_{-1} + w_{\mu} (k) \alpha^{\mu}_{-1} {\tilde c}_{-1} \right]  | \Omega \rangle \! \rangle
 \label{physstate}
\end{eqnarray}

The study of the cohomology of the ${\widetilde Q}_{B}$ operator is very similar, and we get for the tilde string field

\begin{eqnarray}
| {\widetilde \Phi} \rangle \!\rangle &=& \int \frac{d^{26}k}{(2 \pi)^{26}} \left[ {\tilde t}(k) +  {\tilde B}_{\mu}(k) \alpha^{\mu}_{-1} - {\tilde \zeta}(k) c_{-1} - {\tilde \xi} (k) b_{-1} + {\tilde A}_{\mu}(k) {\tilde \alpha}^{\mu}_{-1}\right. \nonumber \\
 &+& \left. {\tilde C}_{\nu \mu}(k) \alpha^{\mu}_{-1} {\tilde \alpha}^{\nu}_{-1} - {\tilde v}_{\mu} (k) b_{-1} {\tilde \alpha}^{\mu}_{-1} - {\tilde w}_{\mu} (k) c_{-1} {\tilde \alpha}^{\mu}_{-1} \right]  | \Omega \rangle \! \rangle
 \label{tildephysstate}
\end{eqnarray}

\section{Computing the Action}

Our extended string field up to the level (1,1) is then

\begin{eqnarray}
| \Phi \rangle \!\rangle &=& \int \frac{d^{26}k}{(2 \pi)^{26}} \left[ t(k) + A_{\mu}(k) \alpha^{\mu}_{-1} + B_{\mu}(k) {\tilde \alpha}^{\mu}_{-1} + \zeta(k) {\tilde c}_{-1} + \xi (k) {\tilde b}_{-1} \right. \nonumber \\
 &+& \left. C_{\mu \nu}(k) \alpha^{\mu}_{-1}{\tilde \alpha}^{\nu}_{-1} + v_{\mu} (k) \alpha^{\mu}_{-1} {\tilde b}_{-1} + w_{\mu} (k) \alpha^{\mu}_{-1} {\tilde c}_{-1} \right]  | \Omega \rangle \! \rangle
\end{eqnarray}

To compute the action, we need the bpz of the field $\langle \! \langle \Phi |$ and $Q_{B} |\Phi \rangle \! \rangle$:

\begin{eqnarray}
\langle \! \langle  \Phi|  &=& \int \frac{d^{26}k}{(2 \pi)^{26}} \langle \! \langle \Omega | \left[ t(-k) + A_{\mu}(-k) \alpha^{\mu}_{1} + B_{\mu}(-k) {\tilde \alpha}^{\mu}_{1} + \zeta(-k) {\tilde c}_{1} - \xi(-k) {\tilde b}_{1} \right. \nonumber \\
 &+& \left. C_{\mu \nu}(-k) \alpha^{\mu}_{1}{\tilde \alpha}^{\nu}_{1} - v_{\mu} (-k) \alpha^{\mu}_{1} {\tilde b}_{1} + w_{\mu} (-k) \alpha^{\mu}_{1} {\tilde c}_{1} \right]
\end{eqnarray}
and

\begin{eqnarray}
&&Q_{B} | \Phi \rangle \! \rangle = \int \frac{d^{26}k}{(2 \pi)^{26}} \left[ t(k) (\alpha' k^{2} - 1) c_{0} + \sqrt{\frac{\alpha'}{2}} k^{\mu} A_{\mu}(k) c_{-1} + (\alpha' k^{2}) A_{\mu}(k) \alpha^{\mu}_{-1} c_{0} \right. \nonumber \\
&&+ (\alpha' k^{2} -1) B_{\mu}(k) {\tilde \alpha}^{\mu}_{-1} c_{0} + (\alpha' k^{2} -1) \zeta(k) c_{0} {\tilde c}_{-1}  + (\alpha' k^{2} -1) \xi(k) c_{0} {\tilde b}_{-1}  \\
&&+ (\alpha' k^{2}) C_{\mu \nu}(k) \alpha^{\mu}_{-1} {\tilde \alpha}^{\nu}_{-1} c_{0} + \sqrt{\frac{\alpha'}{2}} k^{\mu} C_{\mu \nu}(k) {\tilde \alpha}^{\nu}_{-1} c_{-1} + (\alpha' k^{2}) v_{\mu}(k) \alpha^{\mu}_{-1} c_{0} {\tilde b}_{-1} \nonumber \\
&&+ \left. \sqrt{\frac{\alpha'}{2}} k^{\mu} v_{\mu}(k) c_{-1} {\tilde b}_{-1} + (\alpha' k^{2}) w_{\mu}(k) \alpha^{\mu}_{-1} c_{0} {\tilde c}_{-1} + \sqrt{\frac{\alpha'}{2}} k^{\mu} w_{\mu}(k) c_{-1} {\tilde c}_{-1} \right] | \Omega \rangle \! \rangle \nonumber
\end{eqnarray}

Plugging these in (\ref{kinetic-term}), we get\footnote{Where we have used the following prescription
\begin{eqnarray}\label{inner-productk}
 \langle \! \langle \Omega' | c_{0} \tilde{c}_{0} |\Omega \rangle \! \rangle = \langle \tilde{0}, -k' |\langle 0, k'| c_{-1} c_{0} c_{1} \tilde{c}_{-1} \tilde{c}_{0} \tilde{c}_{1} | 0 ; k \rangle  | {\tilde 0}; -k \rangle  \equiv (2 \pi)^{26} \delta^{26}(k-k') \nonumber
\end{eqnarray}}

\begin{eqnarray}
S &=& \frac{\alpha'}{2} \int d^{26}x ~ \left[ \partial^{\mu} t(x) \partial_{\mu} t(x) -\frac{1}{\alpha'} t^{2}(x) + \partial^{\nu} A_{\mu}(x) \partial_{\nu} A^{\mu}(x)   \right. \nonumber \\
&+& \left.  \partial^{\nu} B_{\mu}(x) \partial_{\nu} B^{\mu}(x) -\frac{1}{\alpha'} B_{\mu}(x) B^{\mu}(x) + \partial^{\rho} C_{\mu \nu}(x) \partial_{\rho} C^{\mu \nu}(x) \right]
\end{eqnarray}
Using the decomposition of $C_{\mu \nu}$

\begin{eqnarray}\label{finalaction}
S &=& \frac{\alpha'}{2} \int d^{26}x ~ \left[ \partial^{\mu} t(x) \partial_{\mu} t(x) -\frac{1}{\alpha'} t^{2}(x) + \partial^{\nu} A_{\mu}(x) \partial_{\nu} A^{\mu}(x)   \right. \nonumber \\
&+&  \partial^{\nu} B_{\mu}(x) \partial_{\nu} B^{\mu}(x) -\frac{1}{\alpha'} B_{\mu}(x) B^{\mu}(x) + \partial^{\rho} g_{\mu \nu}(x) \partial_{\rho} g^{\mu \nu}(x)  \\
&+&  \left. \partial^{\rho} B_{\mu \nu}(x) \partial_{\rho} B^{\mu \nu}(x) + \partial^{\rho} \phi(x) \partial_{\rho} \phi(x) \right] \nonumber
\end{eqnarray}

Notice that the fields $v_{\mu} , w_{\mu}, \zeta, \xi $ do not appear in the action, even though they are not ruled out by the cohomology analysis. The reason is that the inner product (\ref{innerprod}) that we have chosen so as to define the free action is the simplest but not the most general one. In fact, by defining the product through the more general insertion
\be\label{innerprod-mod}
{\cal G} = l c_0 + \tilde{l} \tilde{c}_0 + m \, c_{-1} c_0 c_{1} + \tilde{m} \, \tilde{c}_{-1}\tilde{c}_0 \tilde{c}_{1} + n \, b_{-1} c_{0} b_{1} + \tilde{n} \, {\tilde b}_{-1} {\tilde c}_{0} {\tilde b}_{1} + \dots  \, ;
\ee
for $l, \tilde{l}, m, \tilde{m}, n, \tilde{n}  \in \Re$, then all the fields present in the cohomology are kept in the kinetic term of the action

\begin{eqnarray}\label{finalaction-mod}
S &=& \frac{\alpha'}{2} \int d^{26}x ~ \tilde{l}\left[ \partial^{\mu} t(x) \partial_{\mu} t(x) -\frac{1}{\alpha'} t^{2}(x) + \partial^{\nu} A_{\mu}(x) \partial_{\nu} A^{\mu}(x)   \right. \nonumber \\
&+&  \partial^{\nu} B_{\mu}(x) \partial_{\nu} B^{\mu}(x) -\frac{1}{\alpha'} B_{\mu}(x) B^{\mu}(x) + \partial^{\rho} g_{\mu \nu}(x) \partial_{\rho} g^{\mu \nu}(x)  \nonumber \\
&+&  \left. \partial^{\rho} B_{\mu \nu}(x) \partial_{\rho} B^{\mu \nu}(x) + \partial^{\rho} \phi(x) \partial_{\rho} \phi(x) \right] \\
&+&  \tilde{m} \left[ \partial^{\nu} v_{\mu}(x) \partial_{\nu} v^{\mu}(x) + \partial^{\mu} \xi(x) \partial_{\mu} \xi(x) - \frac{1}{\alpha'} \xi^{2} (x) \right] \nonumber \\
&+& {\tilde n} \left[ - \partial^{\nu} w_{\mu}(x) \partial_{\nu} w^{\mu}(x) - \partial^{\mu} \zeta(x) \partial_{\mu} \zeta(x) + \frac{1}{\alpha'} \zeta^{2}(x) \right]   \nonumber
\end{eqnarray}

Therefore, at this point, we can stress three important remarks:

(a) These extra fields can be eliminated from the theory just by taking the appropriate inner product ($m=\tilde{m}= n\dots =0$), and so one recovers the action (\ref{finalaction});

(b) Although the product (\ref{innerprod}) could have some degree of arbitrariness, the insertion (\ref{innerprod-mod}) is the most general one since it allows to capture in the action all the fields physically admissible;

(c) Notice that we need $\tilde{l} , \tilde{m} \geq 0$ and $\tilde{n}\leq 0$ in order to have a positive definite action.

\section{Interaction Terms}

It is not straightforward to give a prescription for the interacting term involving the double string field. However, we would like to finish this paper suggesting how it could be done and leave the checks of detailed S-matrix calculations for a forthcoming paper. The minimal requirement we can do is that the extended three-vertex, contracted with extended fields, should give terms containing the conventional open string interaction terms.

So first, let us observe that the physical field (\ref{physstate}) has the following structure:
\begin{eqnarray}\label{state-decomposition}
| \Phi \rangle \!\rangle &=& | \phi \rangle \otimes | \widetilde{\Omega} \rangle + | \rho \rangle \!\rangle
\end{eqnarray}
where

\begin{eqnarray}
| \phi \rangle  &=& \int \frac{d^{26}k}{(2 \pi)^{26}} \left[ t(k) + A_{\mu}(k) \alpha^{\mu}_{-1}  + \dots \right]  | \Omega \rangle
\end{eqnarray}
is the conventional open-string field, and

\begin{eqnarray}
| \rho \rangle \!\rangle &\equiv& \int \frac{d^{26}k}{(2 \pi)^{26}} \left[ C_{\mu \nu}(k) \alpha^{\mu}_{-1}{\tilde \alpha}^{\nu}_{-1} + B_{\mu}(k) {\tilde \alpha}^{\mu}_{-1} + \zeta(k) {\tilde c}_{-1} + \xi (k) {\tilde b}_{-1} \right. \nonumber \\
 &+& \left.  v_{\mu} (k) \alpha^{\mu}_{-1} {\tilde b}_{-1} + w_{\mu} (k) \alpha^{\mu}_{-1} {\tilde c}_{-1} \right]  | \Omega \rangle \! \rangle\,\,
\end{eqnarray}
wheras the tilde string field has a similar expression:
\begin{eqnarray}\label{state-decomposition-tilde}
| \widetilde{\Phi} \rangle \!\rangle &=& | \Omega \rangle \otimes | \widetilde{\phi} \,\rangle + |\widetilde{ \rho}\, \rangle \!\rangle
\end{eqnarray}

Therefore, in order to satisfy the requirement above, we propose the vertex to be linearly expressible in terms of the conventional one as
\begin{equation}
|V_{\hat{1} \hat{2} \hat{3}} \rangle \!\rangle \equiv |V_{123}\rangle \otimes |K_{\tilde{1}\tilde{2}\tilde{3}}\rangle + |W_{\hat{1} \hat{2} \hat{3}} \rangle \!\rangle
\label{extvertex}
\end{equation}
where
\be
\label{ghostK3}
\widetilde{gh}_{i} [ K] = 2 \,\,\, \forall i=1,2,3.
\ee
so, as for the kinetic term, one shall insert the appropriate ${\cal G}$'s to have a consistent inner product. We have, then, that the extended interaction term in the action (\ref{formal-ext-action}) would read as
\begin{equation}\label{Sint-ext}
 S_{int} =   \frac{g}{3}\langle\!\langle \Phi, \Phi \star \Phi \rangle \!\rangle_{ext} = \frac{g}{3} \langle\!\langle V_{\hat{1} \hat{2} \hat{3}} |\Phi_{1}\rangle \!\rangle|\Phi_{2}\rangle \!\rangle|\Phi_{3}\rangle \!\rangle = \frac{g}{3} \langle V_{123}|\phi_{1}\rangle |\phi_{2}\rangle|\phi_{3}\rangle \left( \langle K_{\tilde{1}\tilde{2}\tilde{3}}| \tilde{\Omega}_{\tilde{1}\tilde{2}\tilde{3}}\rangle\right) +\dots
\end{equation}
where $\dots$ stands for terms involving $|W_{\hat{1} \hat{2} \hat{3}} \rangle \!\rangle$ or $| \rho \rangle \! \rangle$, and

\be
| \tilde{\Omega}_{\tilde{1}\tilde{2}\tilde{3}}\rangle \equiv | \widetilde{\Omega}_{\tilde{1}}\rangle| \tilde{\Omega}_{\tilde{2}}\rangle| \tilde{\Omega}_{\tilde{3}}\rangle
\ee
Thus, the tensor $K$ must have a non-vanishing projection onto this state, and any choice of the other components and $W$ satisfies the requirement of containing the conventional interaction terms. At this point of the present construction, however, there is no additional reasons or properties to give some specific form to these contributions. We can, then, give here the simplest prescription for the vertex extension:
\begin{equation}
|V_{\hat{1} \hat{2} \hat{3}} \rangle \!\rangle \equiv  |V_{123}\rangle \otimes \, {\cal G}_3 \,|{\widetilde \Omega}_{\tilde{1}\tilde{2}\tilde{3}}\rangle \,
\label{extvertex-prescription}
\end{equation}
where $|W_{\hat{1} \hat{2} \hat{3}} \rangle \!\rangle$ is not present, and $|K_{\tilde{1}\tilde{2}\tilde{3}}\rangle \equiv {\cal G}_3 |{\widetilde \Omega}_{\tilde{1}\tilde{2}\tilde{3}}\rangle $ .
The operator ${\cal G}_3 \equiv {\cal G}_{\tilde{1}} {\cal G}_{\tilde{2}} {\cal G}_{\tilde{3}} \sim \tilde{l}_{1} \tilde{l}_{2} \tilde{l}_{3} \tilde{c}^{(\tilde{1})}_{0} \tilde{c}^{(\tilde{2})}_{0} \tilde{c}^{(\tilde{3})}_{0}+ \dots$ is necessary for ghost counting in order to have $\langle K_{\tilde{1}\tilde{2}\tilde{3}}| \tilde{\Omega}_{\tilde{1}\tilde{2}\tilde{3}}\rangle \neq 0$ in (\ref{Sint-ext}). Notice that the vertex defined according to (\ref{extvertex-prescription}) satisfies the known properties of the usual three-string vertex such as cyclicity and BRST-charge conservation.

Finally, by taking the tilde of this object, the complete interaction term of (\ref{formal-ext-action}) is defined as
\begin{equation}
\widehat{S}_{int} \equiv S_{int} - \widetilde{S}_{int} \,,
\end{equation}
where
\begin{equation}
|\widetilde{V}_{\hat{1} \hat{2} \hat{3}} \rangle \!\rangle \equiv |\widetilde{V}_{\tilde{1}\tilde{2}\tilde{3}}\rangle \otimes |{\widetilde K}_{123} \rangle   + |\widetilde{W}_{\hat{1} \hat{2} \hat{3}} \rangle \!\rangle\,\,.
\label{extvertextilde}
\end{equation}
According to the prescription (\ref{extvertex-prescription}), we have $|\widetilde{W}_{\hat{1} \hat{2} \hat{3}} \rangle \!\rangle = 0$.

One first check of our prescription is, for instance, the computation of transition amplitudes involving one graviton state (e.g decaying into fotons/taquions, etcetera). To do this one shall compute (\ref{Sint-ext}) where $|\Phi_{1}\rangle \!\rangle \equiv C_{\mu\nu}\,\a^\mu_{-1}\widetilde{\a}^\nu_{-1}|\Omega_1\rangle \!\rangle$, using (\ref{extvertex-prescription}):

\be
\frac{g}{3} \langle\!\langle V_{\hat{1} \hat{2} \hat{3}} |\left(C^{}_{\mu\nu}\a^\mu_{-1}\widetilde{\a}^\nu_{-1}|\Omega\rangle \!\rangle\right)|\Phi_{2}\rangle \!\rangle|\Phi_{3}\rangle \!\rangle = \frac{g}{3} \langle V_{123}| \langle{\widetilde \Omega}_{\tilde{1}\tilde{2}\tilde{3}}| {\cal G}^{bpz}_3 |\left(C^{}_{\mu\nu}\a^\mu_{-1}\widetilde{\a}^\nu_{-1}|\Omega_1\rangle \!\rangle\right)|\Phi_{2}\rangle \!\rangle|\Phi_{3}\rangle \!\rangle \nonumber
\ee
\be
= \frac{g}{3} \langle V_{123}| \langle{\widetilde \Omega}_{\tilde{2}\tilde{3}}|\left(\langle{\widetilde \Omega}_{\tilde{1}}|
{\cal G}^{bpz}_3 C^{}_{\mu\nu}\a^\mu_{-1}\widetilde{\a}^\nu_{-1}|\Omega_1\rangle \!\rangle\right)|\Phi_{2}\rangle \!\rangle|\Phi_{3}\rangle \!\rangle =0\,\,\,
\ee
which shows that these amplitudes vanish identically at order $g$, so the first non-trivial contribution for these type of processes is order $g^2$, as one should expect from computations of amplitudes with first quantized closed strings.
More detailed consequences of this recipe on computations will be exhaustively explored elsewhere.

\subsection{Conventional OSFT}

The extended action constructed here reproduces the conventional OSFT states and dynamics at zero temperature. Let us briefly show how it is recovered.

As often pointed out along the work, the interpretation of the extended field is statistical and many fields are interpreted as degrees of freedom that emerge from entanglement.
Observe that the decomposition of a general state (at zero temperature) (\ref{state-decomposition}) is \emph{unique}, so clearly, if this state is disentangled (pure) then the rest $|\rho \rangle \!\rangle$ must be expressed as $| \rho' \rangle \otimes | \widetilde{\Omega} \rangle $ in which case $\rho'$ would be absorbed into $|\phi \rangle$. Therefore, by demanding the purity of the states \footnote{In the final discussion, we will propose a suitable definition of entropy that vanishes for pure (disentangled) ones.}, we have $\rho=0$.

The field content of this sector is the conventional open string field one, and remarkably, by virtue of our prescription for the extended vertex (\ref{extvertex-prescription}), this condition is dynamically preserved. One can see this straightforwardly by computing the S-matrix at tree level, using (\ref{extvertex-prescription}), for initial states with $\rho_1=\rho_2=0$, and see that it always gives \emph{out}-states such that $\rho_3=0$. The usual TFD-double sector (at zero temperature) is given by $\tilde{\rho} = 0$ at eq. (\ref{state-decomposition-tilde}).

\section{Conclusions and final remarks}

In the present work, the application of the TFD rules to OSFT was revisited owing to the more current covariant formulation.

We must stress that the present approach differs radically from Leblanc's study where the TFD duplication is realized conventionally on the component fields of the string field and the final results are not substantially different from TFD on conventional quantum field theory, giving the thermal green functions for an infinite collection of interacting fields.
  In contrast, the present formulation is based on the TFD duplication of the algebras (even at classical level), and then, of the basis for the string wave functionals.
The immediate result is the new structure and spectrum of fields, different from the conventional OSFT one. It also brought some technical issues that were solved in the paper, namely, to extend the inner product and the star product to the new algebra in order to write down the extended string field action.

In addition, it was found the more general form for that inner product in the extended star algebra, since it has also been shown that such extension controls which fields of the physical spectrum could appear in the action.
The resulting physical spectrum consists of the standard component fields of OSFT and their corresponding TFD-doubles, and, for instance, in the lower energy levels there are spin-one taquions, scalars, and gauge bosons; and it has been also noticed the presence of particles/fields of the spectrum of a closed string. And, as mentioned before, many sectors can be eliminated from the theory by conveniently choosing the inner product.

We have seen that the theory that results from the application of the TFD rules, updated to string-fields (wave functionals of two-strings), contains all the following sectors in its spectrum (at zero temperature):
(i) The conventional open string field (first term, eq (8.1));
(ii) Its conventional TFD copy \cite{leblanc} (first term of eq (8.4))
(iii) closed strings modes (described in section 5 as the \emph{ground states} of the theory), given by levels: $(0,0), (1,1), (2,2)\dots$, and
(iv) many other fields that in fact do not belong to any of the above sectors such as those appearing in the sectors $(1,2), (2,3), \dots$ . Our interpretation of the non-conventional sectors (iii) and (iv) is, as said before, as emerging effects arising from the entanglement of two open string fields.

This in fact should be expected and is a suggestive result in the sense that
the entanglement between the usual open string and its tilde copy produce emergent effects
which cannot be \emph{seen} in the usual theory. This is the central core of emergent phenomena,
where new degrees of freedom might appear due to collective/entanglement behavior.
There exist examples in the literature where the TFD double is not merely a fictitious system, principally in contexts that involve gravitational degrees of freedom. The most sound work in this sense is due to Israel \cite{israel}, where the thermal properties of a black hole are described by TFD, and the TFD-copy degrees of freedom are identified with the field living behind the event horizon in a black hole. Precisely, these degrees of freedom are causally disconnected from the original system, but they ``collaborate'' with it at quantum level through entanglement (see also \cite{laflame}). More recent references on these points of view are in the context of AdS/CFT, (e.g. \cite{eternal}) where gravitational degrees of freedom are believed to \emph{emerge} from ordinary quantum fields (CFT) and quantum entanglement. As an specific example or result on it, in Ref. \cite{VR} Van Raamsdonk showed that a space time geometry is classically connected  due precisely to the quantum entanglement of two conventional quantum mechanical systems (CFT and precisely, its TFD copy).

The present results suggest a novel possibility in the context of string field theory: that closed string states could be viewed as \emph{non-fundamental}, or more technically, that certain mixed states (backgrounds) of free open strings can be seen as fields of closed string theory. This unifying interpretation is in line with the spirit of the gravity/gauge duality \cite{adscft,bh} and the recent ideas on the spacetime emergence \cite{VR,collapse}, but from a different perspective. Furthermore, the statistical/thermal ingredient of our approach addresses the belief that, at a string field level, the gravitational field should be intimately related to thermodynamical effects \cite{verlinde,macrogeom}.

Although the fact that a closed string can be described in terms of the Hilbert space of two open strings is known, the remarkable result here is that the closed string states can naturally emerge when the TFD rules are properly applied to axiomatic OSFT. On the other hand, the simple need of describing non-pure states of the open string correctly justifies us to formulate this paradigm.

Finally, let us remark that the entropy can be defined straightforwardly in the present framework. We have claimed that the fields of this theory generically represent non-pure states of open strings, so one is able to canonically define the reduced density matrix and the associated entropy.
Given a state $| \Omega(\theta)\rangle\! \rangle$, the reduced density matrix defines as
\be
\rho \equiv \widetilde{Tr} | \Omega(\theta)\rangle\! \rangle   \langle\! \langle \Omega(\theta)|
\ee
where $ \widetilde{Tr}$ denotes trace on the tilde basis elements. So the TFD entropy operator \cite{ume2}, or modular hamiltonian \cite{haag}, can be defined as $K \equiv - \log \,\rho$, and so the entropy of the state is nothing but  $S \equiv Tr \rho \,K \,=\,- Tr \rho \,\log\, \rho$, which is not easy to compute for generic states.  Observe that although, in principle, this can be defined for any string field $|\psi\rangle\!\rangle$,
  only for ground states, $\rho (\Omega) = \tilde{\rho} (\Omega) $ and then $S= \tilde{S}$, as usual for entanglement entropy. Moreover, it is also noticeable that the simplest states with non-vanishing entropy are such that contain graviton-like fields, since the contributions to $K$ come from algebraic combinations as $\alpha^{\mu}_{-1}{\tilde \alpha}^{\nu}_{-1}$ (so as the higher level ones  $\alpha^{\mu}_{-n}{\tilde \alpha}^{\nu}_{-n}$).

This article is an initial study devoted to set the basic structure of the theory, and the thermal effects shall even be introduced through the KMS condition (axiom (iii)) or by minimizing some consistent definition of free energy \cite{ume4,ume1}. In forthcoming works, we shall investigate this and the notion of Bogoliubov transformations $G$ connecting different thermal vacua, which should preserve the algebra of constraints of first quantized strings, namely $[Q_B , G]=0$ \cite{ojima, emch}. In summary, the present approach can be seen as an extension of OSFT, \`a la TFD in the way described above, independent of the fact that thermal equilibrium and temperature have not been introduced yet.

Another issue that needs to be addressed in the future is the study of the solitonic solutions of the equations of motion of the extended theory (Section 8). Trivial extensions can be built from the unextended theory as $| \phi_{sol} \rangle \otimes | \widetilde{\Omega} \rangle$, where $| \phi_{sol} \rangle$ refers to the known solutions of the conventional OSFT \cite{RSZ3,RSZ2,Schn2,RZ,EScn,KORZ,KOSoler,BMTolla,BGTolla1,EMac,EKroyter}. There could be, however, more solutions of the extended theory, which probably involve excitations of  the gravitational field.

\section{Acknowledgements}

 The authors are grateful to CBPF for the warm hospitality during the accomplishment of this paper. Special thanks are due to J. A. Helay\"el for discussions and useful comments.
MBC was partially supported by the grants CONICET PIP 0595/13 and PCI/BEV-MCTI.


\end{document}